\documentclass[aps,prl,twocolumn]{revtex4}

\usepackage{graphicx}
\usepackage{dcolumn}
\usepackage{bm}
\usepackage{amsmath}
\usepackage{amssymb}
\usepackage{amsfonts}
\usepackage{amsthm}
\usepackage{graphicx}
\usepackage{hyperref}
\usepackage{color}
\usepackage{array} 
\usepackage{multirow}
\usepackage{subfigure}
\usepackage{accents}
\usepackage{bbm}
\usepackage{appendix}

\newcommand{\abs}[1]{\left\lvert #1 \right\rvert}

\begin{document}

\title{Super-stable tomography of any linear optical device}

\author{Anthony Laing}
\email{anthony.laing@bristol.ac.uk}
\author{Jeremy L. O'Brien}
\affiliation{Centre for Quantum Photonics, H. H. Wills Physics Laboratory \& Department of Electrical and Electronic Engineering, University of Bristol, BS8 1UB, United Kingdom}

\begin{abstract}
Linear optical circuits of growing complexity are playing an increasing role in emerging photonic quantum technologies. Individual photonic devices are typically described by a unitary matrix containing amplitude and phase information, the characterisation of which is a key task.  We present a constructive scheme to retrieve the unitary matrix describing an arbitrary linear optical device using data obtained from one-photon and two-photon ensembles.  The scheme is stable on the arbitrarily increasable length scale of the photon packet and independent of photon loss at input and output ports of the device. We find a one-to-one correspondence between ideal data and unitary matrix, and identify the class of non-unitary matrices capable of reproducing the data.  The method is extended for coherent state probes, which can simulate two-photon statistics with a reduced visibility.  We analyse the performance of reconstruction to simulated noise.
\end{abstract}

\maketitle
Photonic circuitry is set to play an important role in emerging quantum technologies \cite{nielsenchuang}, connecting or implementing quantum logic gates \cite{KLM, ob-nat-426-264}, enabling photonic simulators \cite{pe-sci-329-1500, aaronsonQIP2010}, and coding states for quantum communication \cite{ce-prl-88-127902, rfiQKD}, with integrated optics \cite{po-sci-320-646} enabling large reconfigurable circuits \cite{ma-nphot-3-346, sh-nphot-6-45, bo-prl-108-053601}.  In these contexts, each lossless linear optical network is ideally described by a unitary matrix \cite{reck}, the determination of which is a key task.  While a general quantum optical process may be fully characterised with coherent light and homodyne detection \cite{lo-sci-322-563}, the device to be determined and the probing set-up must be stable with respect to one another on a sub-wavelength scale.

Since the phases imparted to each photon of a two-photon state combine to a single global phase, circuit probing via two-photon interferometry \cite{hom, ma-apb-60-s111, pe-natcom-2-224, la-prl-108-260505} is stable on the scale of the photon packet, which is typically orders of magnitude greater than wavelength.  Photons generated from spontaneous parametric down conversion (SPDC) can be lengthened through spectral filtering or with a cavity in narrowband SPDC \cite{ha-ol-34-55}, while photons generated in an atom-cavity system can be many meters long \cite{le-apb-77-797}.  Two-photon sources are commonplace in quantum optics labs, with quantum interference visibilities approaching unity \cite{la-apl-97-211109}, and future many-photon experiments are likely to be easily configured to produce subsets of data from one-photon and two-photon ensembles for characterisation purposes.  Additionally, two-photon quantum statistics can be classically simulated by two coherent beams with a mutually randomised phase \cite{br-prl-102-253904, ke-pra-81-23834} so that any circuit information retrievable with a two-photon state is also available to classical states.

In Ref. \cite{pe-natcom-2-224} the transfer matrix of a four-mode linear optical device was numerically approximated as the matrix that minimised the distance between a self-simulated data-set of all possible two-photon statistics and experimental data, with data from one-photon ensembles used as starting parameters. However, a numerical search among the space of $m \times m$ dimensional unitary matrices based on the $\approx m^{4}$ data elements associated with all possible two-photon statistics becomes challenging for large systems.  More fundamentally, it has not been clear up to this point if data from one-photon ensembles (one-photon data) and data two-photon ensembles (two-photon data) are sufficient to uniquely determine the matrix description of a linear optical circuit.

\begin{figure}[t]
\vspace{5 pt}
\includegraphics[trim=0 0 0 0, clip,width=.975\columnwidth]{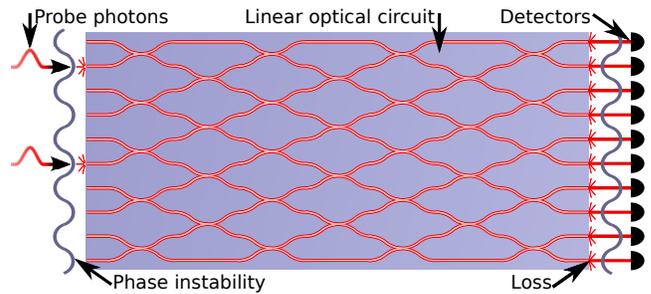}
\vspace{-10 pt}
\caption{
Circuit tomography with one-photon and two-photon probes is insensitive to phase fluctuations and independent of loss at input and output ports. }
\label{fgSchematic}
\vspace{-15 pt}
\end{figure}

Here, we give a constructive method to efficiently determine the set of matrices that produce a given set of ideal one-photon and two-photon data, for an arbitrary linear optical network.  Reconstruction is insensitive to fluctuating phases between input ports and between output ports, as shown in Fig~\ref{fgSchematic}, and therefore does not require active stabilisation or other techniques for stable classical interferometry.  In addition, reconstruction is independent of the typically unknown and tomographically troublesome photon losses at device input and output ports.  The scheme involves injecting one-photon and two-photon states into the linear optical network with correlated photon detection, as shown in Fig \ref{fgSchematic}.  Matrix amplitude information is probed with one-photon ensembles, while specific phase parameters are targeted by specific two-photon ensembles, so that the required number of measurements scales quadratically with the number of modes, and linearly with the number of unknown parameters.  Assuming the device can be described by a unitary matrix, applying $2m$ of the $m^{2}$ orthonormal constraints leads to a one-to-one mapping between idealised experimental data and unitary matrix, up to the fluctuating phases.  If the port-losses are known to be zero, then a matrix can be recovered with no orthonormal constraints invoked or required.  We consider non-ideal situations and simulate noisy data from circuits of up to $20$ optical modes to quantify the fidelity of reconstruction versus noise.

\begin{figure}[t]
\vspace{0 pt}
\includegraphics[trim=0 0 0 0, clip,width=0.95\columnwidth]{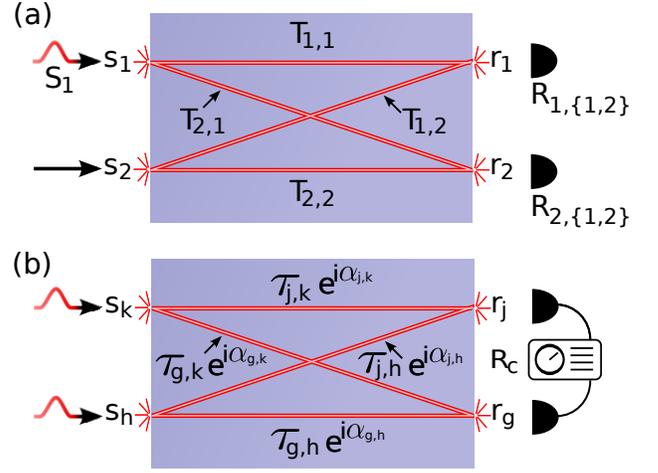}
\vspace{-5 pt}
\caption{Probing of circuit sub-section.
(a) One-photon probing of a two-mode subsection of larger device.
(b) Two-photon targeting of phase parameters, with $R_{c}$ indicating correlated detection.
}
\label{fg2x2}
\vspace{-10 pt}
\end{figure}

An insightful example of the tools used in our method is the case of a two-mode device, or subsection of a larger device, as shown in Fig~\ref{fg2x2}(a).  $S_{k}$ photons are individually injected into input port $k$, coupled with efficiency $s_{k}$, transmitted to output port $j$ with probability $T_{j,k}$, and detected with efficiency $r_{j}$, giving a count rate of $R_{j,k}$
\footnote{We assume detection on only the indicated mode and any hidden modes, e.g. polarisation, are filtered out.}.
With $R_{j,k} = S_{k} \; s_{k} \; T_{j,k} \; r_{j}$, four such measurements combine to constrain four transition probabilities $\{T_{j,k}\}$ independently of coupling efficiencies: 
\begin{align}
\frac{R_{1,1}}{R_{1,2}} = \frac{S_{1}}{S_{2}} \; \frac{s_{1}}{s_{2}} \; \frac{T_{1,1}}{T_{1,2}}, & \;\;\;\;\;\;\;\; \frac{R_{2,2}}{R_{2,1}} = \frac{S_{2}}{S_{1}} \; \frac{s_{2}}{s_{1}} \; \frac{T_{2,2}}{T_{2,1}} \nonumber
\\[6pt]
\implies \frac{T_{1,1} \, T_{2,2}}{T_{1,2} \, T_{2,1}} &= \frac{R_{1,1} \, R_{2,2}}{R_{1,2} \, R_{2,1}}.
\label{eq2x2}
\end{align}
In the case that the $2\times2$ device behaves unitarily, probability constraints (e.g. $T_{1,1}=T_{2,2}=1-T_{1,2}=1-T_{2,1}$) allow description in terms of one parameter,
\begin{equation}
\frac{T^{2}_{1,1}}{(1-T_{1,1})^{2}}=\frac{R_{1,1} \, R_{2,2}}{R_{1,2} \, R_{2,1}}, \nonumber
\label{eq2x2consApp}
\end{equation}
to derive the experimentalist's loss-insensitive equation for the reflectivity of a beamsplitter, $T_{1,1} \equiv B_{R}$,
\begin{equation}
B_{R}=\frac{\sqrt{X}}{1+\sqrt{X}}, \nonumber
\label{eq2x2BS}
\end{equation}
where $X$ replaces the RHS of \eqref{eq2x2}.

An alternative way to find the reflectivity of a beamsplitter uses the visibility of quantum versus classical correlations in a Hong Ou Mandel (HOM) dip \cite{hom}.  More generally, the visibility of a HOM dip can be used to apply loss-insensitive constraints to the $2\times2$ submatrix $S_{2}$ that describes a two mode system, which may be part of a larger device.  $S_{2}$ consists of real positive probability amplitudes $\tau$ and unit complex phase factors $e^{i \alpha}$; pre and post multiplication by diagonal matrices $D_{c}$ and $D_{d}$ includes coupling and detection efficiencies, to describe of the overall subprocess $P_{2} = D_{d} \, S_{2} \, D_{c}$,
\begin{equation*}
P_{2} =
\begin{pmatrix}
\sqrt{r_{j}} & 0\\
0 & \sqrt{r_{g}}
\end{pmatrix}
\begin{pmatrix}
\tau_{j,k} e^{i\alpha_{j,k}} & \tau_{j,h} e^{i\alpha_{j,h}}\\
\tau_{g,k} e^{i\alpha_{g,k}} & \tau_{g,h} e^{i\alpha_{g,h}}
\end{pmatrix}
\begin{pmatrix}
\sqrt{s_{k}} & 0\\
0 & \sqrt{s_{h}}
\end{pmatrix}
\label{eqm}
\end{equation*}
as shown in Fig~\ref{fg2x2}(b).

Simultaneously injecting two photons into the $P_{2}$ device and counting subsequent coincidental detector clicks as a proportion, gives a quantum signal equal to the square of the absolute value of the permanent of $P_{2}$: 
\begin{align}
Q_{g,h,j,k}
& = \abs{ Per\left( P_{2} \right)}^{2} \nonumber \\
& = s_{h} \, s_{k} \, r_{g} \, r_{j}  \left(\tau^{2}_{j,k}  \; \tau^{2}_{g,h} \; + \; \tau^{2}_{g,k} \; \tau^{2}_{j,h} \right) \nonumber \\
& + s_{h} \, s_{k} \, r_{g} \, r_{j} \;\; \tau_{j,k}  \; \tau_{j,h} \; \tau_{g,k} \; \tau_{g,h} \nonumber \\
& \times 2 \cos{(\alpha_{j,k} - \alpha_{j,h} - \alpha_{g,k} + \alpha_{g,h})}.
\label{eqQ}
\end{align}
Introducing a delay between the photons so that no quantum interference occurs gives the classical signal equal to the square of the permanent of the matrix containing absolute values of the entries in $P_{2}$:
\begin{eqnarray}
C_{g,h,j,k}
&=& Per\left(   \abs{P_{2}}^{2}   \right) \nonumber \\[6pt]
&=& s_{h} \, s_{k} \, r_{g} \, r_{j} \left(\tau^{2}_{j,k}  \; \tau^{2}_{g,h} \; + \; \tau^{2}_{g,k} \; \tau^{2}_{j,h} \right) \nonumber \\[6pt]
&=& R_{j,k}  \; R_{g,h} \; + \; R_{g,k} \; R_{j,h}
\label{eqC}
\end{eqnarray}
and can be calculated directly from one-photon data.  The loss insensitive visibility is calculated as
\begin{eqnarray}
V_{g,h,j,k}
&\equiv& \; \frac{C_{g,h,j,k} - Q_{g,h,j,k}}{C_{g,h,j,k}}\\[6pt]
\label{eqVdef}
&=& \; -2 \cos{(\alpha_{j,k} - \alpha_{j,h} - \alpha_{g,k} + \alpha_{g,h})}\nonumber\\[6pt]
&\times& \; \frac{\; \tau_{j,k}  \; \tau_{j,h} \; \tau_{g,k} \; \tau_{g,h}}{\tau^{2}_{j,k}  \; \tau^{2}_{g,h} \; + \; \tau^{2}_{g,k} \; \tau^{2}_{j,h}}\nonumber.
\label{eqV}
\end{eqnarray}

Probability amplitudes $\tau$ are the square root of the transition probabilities $T$ described in \eqref{eq2x2} so can be constrained from
\begin{align}
X_{g,h,j,k} \equiv \frac{R_{j,k} \, R_{g,h}}{R_{j,h} \, R_{g,k}}, & \;\;\; x_{g,h,j,k} \equiv \sqrt{X_{g,h,j,k}} \nonumber
\\[6pt]
\implies x_{g,h,j,k} &= \frac{\tau_{j,k} \, \tau_{g,h}}{\tau_{j,h} \, \tau_{g,k}}.
\label{eq2x2G}
\end{align}
A loss-independent measurable constraint on $S_{2}$ phase angles then follows from
\begin{align}
y_{g,h,j,k} \equiv x_{g,h,j,k} + x^{-1}_{g,h,j,k} &\implies \nonumber
\\[6pt]
\cos{(\alpha_{j,k} - \alpha_{j,h} - \alpha_{g,k} + \alpha_{g,h})} &= -\frac{V_{g,h,j,k} \; y_{g,h,j,k}}{2}.
\label{eqang}
\end{align}
Since cosine is a symmetric function, the sign of the argument is unknown.  However, the sign for any one of the four phase angles, say $\alpha_{g,h}$, can be ascertained from \eqref{eqang} if its absolute value $0 \leq |\alpha_{g,h}| < \pi$ is known, and the absolute values \emph{and} signs of the other three (reference) angles are known, which will be the case, as we will see.  The sign on $\alpha_{g,h}$ can be found by labelling the argument from \eqref{eqang} as $\beta_{g,h,j,k} \equiv | \alpha_{j,k} - \alpha_{j,h} - \alpha_{g,k} + \alpha_{g,h} |$ then computing
\begin{align}
\operatorname{sgn}&[\alpha_{g,h}] = \operatorname{sgn}[ \nonumber \\
&|\beta_{g,h,j,k} - |\alpha_{j,k} - \alpha_{j,h} - \alpha_{g,k} - |\alpha_{g,h}||| \nonumber \\
-
&|\beta_{g,h,j,k} - |\alpha_{j,k} - \alpha_{j,h} - \alpha_{g,k} + |\alpha_{g,h}||| \;\; ]
\label{eqSign}
\end{align}
assuming the 3 reference angles do not sum to zero
\footnote{In general there will be many ($\approx m^{2}$) choices for a $2\times2$ submatrix that contains $\alpha_{g,h}$.}.

Using the amplitude and phase constraints of \eqref{eq2x2G} and \eqref{eqang} with the sign disambiguation of \eqref{eqSign}, we can reconstruct $M$, the $m$ dimensional matrix description of an arbitrary $m$ mode optical network, up to $2m-1$ of the $\approx m^{2}$ parameters, independently of any unknown losses on input and output ports.  When $M$ can be assumed to behave as a unitary transformation, constraints give a tractable solution system for the $2m-1$ unknown parameters.

We can specify some equivalence among matrices where corresponding linear optical networks within the equivalence class produce the same data.  Firstly, two matrices $M_{a}$ and $M_{b}$ are equivalent if there exist two diagonal unitary matrices $D^{U}_{1}$ and $D^{U}_{2}$ such that $M_{a}=D^{U}_{1}M_{b}D^{U}_{2}\nonumber$.  In a linear optical network $M$, these diagonal matrices can be regarded as the typically unknown and somewhat trivial phases on the input and output ports of the network, to which our data is insensitive, as long as photons are input in a Fock state (not superposed across modes) and there is no post-$M$ interferometry before detection.  These trivial phases may equally be regarded as a particular instance of the fluctuating phases from Fig \ref{fgSchematic}.  We encapsulate this idea in the standard way by insisting that our $M$ is \emph{real bordered}, that is, the entries on the left-most column and uppermost row should be real and positive: $\{M_{j,1}, M_{1,k}\}^{m}_{j,k=1} \in \mathbbm{R}^{+}$.  Secondly, noting the equality of photon statistics under the complex conjugate transformation $M\rightarrow M^{*}$, we insist that the imaginary part of the element $M_{2,2}$ be non negative: $\operatorname{Im} (M_{2,2}) \geq 0$.

Additionally, we make the following assumptions:  for a straightforward description, we firstly assume that the two leftmost columns and two uppermost rows contain no vanishingly small elements --- if this were not the case, one-photon data would identify zero matrix elements and the reconstruction method can be straightforwardly adapted; secondly, while our method is insensitive to loss on the input and output ports, we assume any losses are not near-total.  With these equivalencies and assumptions, the matrix to be recovered can be parameterised as
\begin{equation}
M_{1}=
\begin{pmatrix}
\tau_{1,1} & \, \tau_{1,2} & \, \cdots & \, \tau_{1,m} \\[8pt] 
\tau_{2,1} & \, \tau_{2,2} \, e^{ i \alpha_{2,2}} & \, \cdots & \, \tau_{2,m} e^{ i \alpha_{2,m}} \\[8pt]
\vdots & \, \vdots & \, \ddots & \, \vdots \\[8pt]
\tau_{m,1} & \, \tau_{m,2} e^{ i \alpha_{m,2}} & \, \cdots & \, \tau_{m,m} e^{ i \alpha_{m,m}}
\end{pmatrix}
\label{eqM1}
\end{equation}
where, as before, the $\tau$ correspond to transition amplitudes and the $\alpha$ correspond to phase angles.

Replacements are made on all elements of $M_{1}$ \emph{apart from} the first column and first row.  The $\tau$ amplitudes are replaced following \eqref{eq2x2G} fixing $\tau_{j,k} = \tau_{1,1}$ while the indices of $\tau_{g,h}$ are taken from the $\tau$ to be replaced.  All $\tau$ not in the first column or row, become functions of the $\tau$ in the first row and column.  Phase angles $-\pi \leq \alpha < \pi$ are found (up to a sign) following \eqref{eqang} with the same index pattern used for the $\tau$, so that only one angle appears in the cosine.

Apart from the sign on $\alpha_{2,2}$, which is defined to be positive, all signs on the $\alpha$ in $M_{1}$ are determined using \eqref{eqSign}.  Signs on the $\alpha$ in the 2nd column are found by fixing $\alpha_{j,k}=\alpha_{2,1}$ while the indices of $\alpha_{g,h}=\{\alpha_{3 ... m, 2}\}$, moving from third row to row $m$.  Similarly, signs on the 2nd row are found by fixing $\alpha_{j,k}=\alpha_{1,2}$ and $\alpha_{g,h}=\{\alpha_{2,3 ... m}\}$, moving from third column to column $m$.  Signs on the remaining $\alpha$ are found by fixing $\alpha_{j,k}=\alpha_{2,2}$ and the indices of $\alpha_{g,h}$ are taken from the $\alpha$ to be replaced.  This pattern ensures that signs on all references phases in \eqref{eqSign} are known, as is required.   With these replacements, the transfer matrix can be parameterised by $2m-1$ unknown border amplitudes
\begin{equation}
M_{2}=\nonumber\\
\begin{pmatrix}
\tau_{1,1} & \, \tau_{1,2} & \, \cdots & \, \tau_{1,m} \\[8pt] 
\tau_{2,1} & \, \tilde{x}_{2,2,1,1} e^{ i \tilde{\alpha}_{2,2}} & \, \cdots & \, \tilde{x}_{2,m,1,1} e^{ i \tilde{\alpha}_{2,m}} \\[8pt]
\vdots & \, \vdots & \, \ddots & \, \vdots \\[8pt]
\tau_{m,1} & \, \tilde{x}_{m,2,1,1} e^{ i \tilde{\alpha}_{m,2}} & \, \cdots & \, \tilde{x}_{m,m,1,1} e^{ i \tilde{\alpha}_{m,m}}
\end{pmatrix}.
\label{eqM2}
\end{equation}
where, for ease of notation, we have introduced
\begin{align}
\tilde{x}_{g,h,j,k} & = x_{g,h,j,k} \, \frac{\tau_{j,h}\tau_{g,k}}{\tau_{j,k}} ,\nonumber
\\[6pt]
\tilde{\alpha}_{g,h}&  = \operatorname{sgn}[\alpha_{g,h}] \arccos (\frac{1}{2} \;V_{g,h,j,k} \; y_{g,h,j,k})\nonumber, 
\label{eqAngLabel}
\end{align}
highlighting that the $\alpha$ are fully determined from experimental measurements and not functions of the $\tau$.

The matrix $E$, describing the complete experimental set up comprising the device $M$ and losses on input and output ports, has transition probabilities corresponding to the set of one photon count rates $R$, while phase angles are the calculated $\alpha$: $E$ is therefore fully characterised so that the output state can be calculated for any arbitrary photon number Fock state input.  This implies that probing the system with more than two photons, or using number resolving detectors, provides no extra information to solve the remaining $2m-1$ unknown parameters in $M_{2}$.  In the case of no losses $M=E$ and full reconstruction has been achieved.  However, zero losses on input and output ports is atypical, particularly when one considers collection efficiencies from a photon source and detection efficiencies.

When $M$ can be assumed to behave unitarily, normalisation can be applied to the first column of $M_{2}$ and orthogonality constraints applied between the first column and the remaining $m-1$ columns, with the same applying for rows.  The solution system is the matrix of coefficients in $M_{2}$,
\begin{equation}
M_{\mu} \equiv
\begin{pmatrix}
1 & \, 1 & \, \cdots & \, 1 \\[8pt] 
1 & \, x_{2,2,1,1} \, e^{ i \tilde{\alpha}_{2,2}} & \, \cdots & \, x_{2,m,1,1} e^{ i \tilde{\alpha}_{2,m}} \\[8pt]
\vdots & \, \vdots & \, \ddots & \, \vdots \\[8pt]
1 & \, x_{m,2,1,1} e^{ i \tilde{\alpha}_{m,2}} & \, \cdots & \, x_{m,m,1,1} e^{ i \tilde{\alpha}_{m,m}}
\end{pmatrix},\nonumber
\label{eqSolve1}
\end{equation}
applied to the transition probabilities, to give the orthonormal constraints,
\begin{equation}
M^{\dagger}_{\mu}
\begin{pmatrix}
\tau^{2}_{1,1}\\[8pt]
\tau^{2}_{2,1}\\[8pt]
\vdots \\[8pt]
\tau^{2}_{m,1}
\end{pmatrix}
=
\begin{pmatrix}
1\\[8pt]
0\\[8pt]
\vdots \\[8pt]
0
\end{pmatrix},
\;\;\;\;\;\;
M_{\mu}
\begin{pmatrix}
\tau^{2}_{1,1}\\[8pt]
\tau^{2}_{1,2}\\[8pt]
\vdots \\[8pt]
\tau^{2}_{1,m}
\end{pmatrix}
=
\begin{pmatrix}
1\\[8pt]
0\\[8pt]
\vdots \\[8pt]
0
\end{pmatrix}.
\label{eqSolve1}
\end{equation}
A standard method for solving systems of linear equations may then be used \footnote{We provide a four mode example for a Haar random unitary in the appendix.}.

One-photon and two-photon statistics can be estimated using classical coherent states.  The classical analogue of one-photon counts $R_{j,k}$ are intensities $I_{j,k}$, measured at port $j$ from injecting a coherent beam of intensity $I_{0}$ into port $k$: the one-photon counts in \eqref{eq2x2} can be replaced by a ratio of classical intensities.  The classical counterpart of Hong Ou Mandel correlations are those of Hanbury-Brown and Twiss \cite{HBT01} arising from interference between two independent fields \cite{ma-pra-28-929, pa-rmp-58-209}, which have already been exploited in linear optical devices to simulate two-photon statistics \cite{br-prl-102-253904, ke-pra-81-23834}.  The quantum coincidence rate $Q_{g,h,j,k}$ can be accurately estimated by injecting two coherent states with the same mean photon number and a randomised relative phase into respective input ports $h$ and $k$.  The intensity correlation function, $\Gamma_{g,h,j,k}$, measured at output ports $g$ and $j$ is
\begin{equation}
\Gamma_{g,h,j,k}=I^{2}_{0}\abs{ Per\left( P_{2} \right)}^{2} + I_{j,k}I_{g,k} + I_{j,h}I_{g,h}
\label{eqGamma1}
\end{equation}
so that the quantum correlation can be estimated from $\Gamma_{g,h,j,k}$ minus the product of individual intensities
\begin{equation}
I^{2}_{0} Q_{g,h,j,k} = \Gamma_{g,h,j,k} - I_{j,k}I_{g,k} - I_{j,h}I_{g,h}.
\label{eqGamma2}
\end{equation}

Experimental noise and device imperfections will drive the reconstructed matrix away from that which best describes the optical device, and away from unitarity since we use only some of those constraints when solving for border amplitudes.  The unitary matrix $U_{p}$ closest to the reconstructed matrix may be found from the polar decomposition \cite{fa-pams-6-111}, but in the presence of errors $U_{p}$ will not, in general, perfectly match any true underlying unitary matrix $U_{t}$.

\begin{figure}[t]
\vspace{0 pt}
\includegraphics[trim=0 35 0 35, clip,width=1\columnwidth]{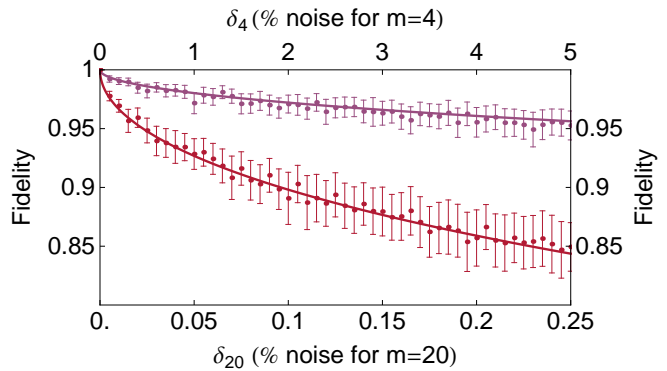}
\vspace{-20 pt}
\caption{Average fidelities of reconstructed unitary matrices versus simulated noise for m=4 (top curve) and m=20 (bottom curve).}
\label{fgFidNoise}
\vspace{-15 pt}
\end{figure}

We analysed the effect of noisy data on our reconstruction procedure, for devices from dimension $m=4$ to $m=20$, over a range of noise level $\delta$.  Noise was added to the simulated experimental values for one-photon count rates $R$ and quantum interference visibilities $V$: the ideal values were multiplied by $(1+\epsilon)$, where $\epsilon$ was chosen from a normal distribution centred at $0$.  The amount of noise $\delta$ was controlled by varying the width of the distribution with $\delta=3\sigma$, the third standard deviation, making it very unlikely that experimental values are found to be much further than $\delta$ away from their ideal values.  For a particular value of $m$ and $\delta$, we generated $1000$ Haar random unitary matrices, each of which was pre and post multiplied by diagonal matrices with real random values to simulate loss.  The algorithm was ran for each instance, with a unitary matrix extracted via the polar decomposition.  The average trace distance was computed over all $1000$ examples to give an average fidelity of reconstruction; the variance of each $1000$ example set was also computed.

Over the tested range of matrix size $m$, we found the empirical relation between average fidelity of reconstruction $F$ and error level $\delta$
\begin{equation}
F\approx \exp{(-\lambda \delta^{\frac{1}{2}})},
\label{eqF}
\end{equation}
with $\lambda \approx \frac{m-3}{5}$ giving good predictions for fidelities down to at least $\approx 85\%$.  Figure \ref{fgFidNoise} shows $51$ simulated data points between $\delta=0\%$ and $\delta=5\%$ for $m=4$, and between $\delta=0\%$ and $\delta=0.25\%$ for $m=20$, with error bars on each point corresponding to the variance of its $1000$ example data set.  The empirical curve $F$ is shown as the solid line.  

The effect of errors analysed here involved no attempt at error correction.  A useful line of research would be to investigate how the impact of errors may be minimised by advantageously permuting the matrix prior to reconstruction.  Additionally, having used only $\approx 2m^{2}$ two-photon measurements from the possible $\approx m^{4}$, it would be interesting to understand how extra subsets of two-photon data may increase robustness to noise, or whether resources should be directed toward reducing errors on the minimal set.

We have presented a scheme in which the unitary matrix of a linear optical device may be constructively retrieved without the requirement for stable classical interferometry.  The merits of probing with coherent states versus single photon states is interesting.  Since the coincidental signal falls off quadratically (due to the number of ways of choosing 2 from $m$ modes) an arbitrarily bright coherent state may seem preferable.  However, two-photon states readily available within a larger multi-photon experiment may exhibit the same characteristics as those producing the larger data set, with statistics more representative of a device that is somewhat defined with respect to its photons; for example, the reflectivity of a waveguide directional coupler is a function of wavelength.  Any advantage from an increase in signal and greater measurement precision with classical states should be weighed against potential inaccuracies if the crucial properties of those states do not match the properties of the photons, due to some experimental uncertainty.  Furthermore, the capability to re-take tomographic data as a subset of principal experimental data, for a freshly reconfigured system, during the course of an uninterrupted experiment, is appealing.

\begin{acknowledgments}
We thank I. Barr, J. Carolan, M. Lobino, E. Mart\'{i}n L\'{o}pez, J. Matthews, N. Russell, and T. Rudolph for helpful discussions.  This work was supported by EPSRC, ERC, and NSQI. J.L.O'B. acknowledges a Royal Society Wolfson Merit Award. 
\end{acknowledgments}

\appendix
\begin{widetext}
\vspace{10 pt}
\section{Appendix 1: A four mode example}
We present a four mode example of simulated data $E^{(sim)}$ for a system comprised of a device described by a Haar random unitary $U^{(sim)}$ with losses on inputs $L^{(in)}$ and outputs $L^{(out)}$

\begin{align*}
E^{(sim)} &= L^{(out)}U^{(sim)}L^{(in)}\\[8pt]
&=
\begin{pmatrix}
 0.46 & 0. & 0. & 0. \\
 0. & 0.65 & 0. & 0. \\
 0. & 0. & 0.41 & 0. \\
 0. & 0. & 0. & 0.37
\end{pmatrix}
\begin{pmatrix}
 0.245 & 0.54 & 0.537 & 0.601 \\
 0.492 & 0.377 +0.192 i & -0.634+0.213 i & 0.027 -0.362 i \\
 0.662 & 0.007 +0.119 i & 0.305 -0.339 i & -0.549+0.196 i \\
 0.509 & -0.633-0.34 i & -0.042+0.235 i & 0.398 +0.095 i
\end{pmatrix}
\begin{pmatrix}
 0.08 & 0. & 0. & 0. \\
 0. & 0.82 & 0. & 0. \\
 0. & 0. & 0.55 & 0. \\
 0. & 0. & 0. & 0.24
\end{pmatrix}\\[8pt]
&=
\begin{pmatrix}
 0.009 & 0.204 & 0.136 & 0.066 \\
 0.026 & 0.201 +0.102 i & -0.227+0.076 i & 0.004 -0.057 i \\
 0.022 & 0.002 +0.04 i & 0.069 -0.076 i & -0.054+0.019 i \\
 0.015 & -0.192-0.103 i & -0.009+0.048 i & 0.035 +0.008 i
\end{pmatrix}
\end{align*}

Following \eqref{eqQ} to \eqref{eqSign} with perfectly simulated quantum interference data and power ratio measurements, the unitary parameterised by 7 probability amplitudes is

\begin{align*}
M^{(sim)}_{2}=
\begin{pmatrix}
 \tau_{1,1} & \tau_{1,2} & \tau_{1,3} & \tau_{1,4} \\
 \tau_{2,1} & (0.348 +0.177 i) \frac{\tau_{1,2} \tau_{2,1}}{\tau_{1,1}} & -(0.588 -0.197 i) \frac{\tau_{1,3} \tau_{2,1}}{\tau_{1,1}} & (0.022 -0.300 i) \frac{\tau_{1,4} \tau_{2,1}}{\tau_{1,1}} \\
 \tau_{3,1} & (0.005 +0.082 i) \frac{\tau_{1,2} \tau_{3,1}}{\tau_{1,1}} & (0.211 -0.234 i) \frac{\tau_{1,3} \tau_{3,1}}{\tau_{1,1}} & -(0.338 -0.121 i) \frac{\tau_{1,4} \tau_{3,1}}{\tau_{1,1}} \\
 \tau_{4,1} & -(0.565 +0.304 i) \frac{\tau_{1,2} \tau_{4,1}}{\tau_{1,1}} & -(0.038 -0.211 i) \frac{\tau_{1,3} \tau_{4,1}}{\tau_{1,1}} & (0.319 +0.076 i) \frac{\tau_{1,4} \tau_{4,1}}{\tau_{1,1}}
\end{pmatrix}
\end{align*}

giving the solution systems for the unknown $\tau$ as

\begin{align*}
\begin{pmatrix}
 1 & 1 & 1 & 1 \\
 1 & 0.348 +0.177 i & 0.005 +0.082 i & -0.565-0.304 i \\
 1 & -0.588+0.197 i & 0.211 -0.234 i & -0.038+0.211 i \\
 1 & 0.022 -0.300 i & -0.338+0.121 i & 0.319 +0.076 i
\end{pmatrix}
\begin{pmatrix}
0.245^{2} \\ 0.492^{2} \\ 0.662^{2} \\ 0.509^{2}
\end{pmatrix}
=
\begin{pmatrix}
1 \\ 0 \\ 0 \\ 0
\end{pmatrix}\\[16pt]
\begin{pmatrix}
 1 & 1 & 1 & 1 \\
 1 & 0.348 -0.177 i & -0.588-0.197 i & 0.022 +0.300 i \\
 1 & 0.005 -0.082 i & 0.211 +0.234 i & -0.338-0.121 i \\
 1 & -0.565+0.304 i & -0.038-0.211 i & 0.319 -0.076 i
\end{pmatrix}
\begin{pmatrix}
0.245^{2} \\ 0.540^{2} \\ 0.537^{2} \\ 0.601^{2}
\end{pmatrix}
=
\begin{pmatrix}
1 \\ 0 \\ 0 \\ 0
\end{pmatrix}
\end{align*}
in agreement with $U^{(sim)}$.
\end{widetext}
\end{document}